
\magnification=1200
\pageno=1
\baselineskip=17pt
\hfill{Saclay, T94/078}
\hfill{Orsay, LPTHE 94/58}
\hfill{(June 1994)}
\vskip 48pt
\centerline{{\bf INTERMITTENCY AND EXOTIC CHANNELS}}
\vskip 48pt
\centerline{{\bf A. Bialas}\footnote{$ ^\ast $}{On leave from: Institute of
Physics, Jagellonian
University, Reymonta 4, PL-30 059 Cracow, POLAND}}
\centerline{LPTHE, B\^atiment 211, Universit\'e Paris-Sud}
\centerline{F-91405 Orsay, FRANCE}
\vskip 12pt
\centerline{{\bf R. Peschanski}}
\centerline{CEA, Service de Physique Th\'eorique, CE de Saclay}
\centerline{F-91191 Gif-sur-Yvette Cedex, FRANCE}
\vskip 48pt
\noindent {\bf ABSTRACT}

It is pointed out that accurate measurements of short-range
two-particle correlations in like-charge $ K\pi $ and in $ \pi^ 0\pi^ 0 $
channels should be
very helpful in determining the origin of the \lq\lq intermittency\rq\rq\
phenomenon
observed recently for the like-charge pion pairs.
\vglue 4truecm
{\it Submitted to Phys. Rev. D (Brief Reports)}
\vfill\eject

It is now well established $ [1,2] $ that the correlations between two pions
at
small relative momenta observed in different hadron-induced processes of
multiparticle production are drastically different for the like and unlike
pairs. The like pairs show the so-called \lq\lq intermittency\rq\rq\ $ [3], $
i.e. a strong
(probably power-law) rise of the correlation function at small $ Q^2, $ where
$$ Q^2= \left(p_1-p_2 \right)^2= \left(p_1+p_2 \right)^2-4m^2_\pi \eqno (1) $$
is the square of the difference of particle momenta. This phenomenon finds a
natural explanation in terms of quantum interference between identical
particles (HBT correlations $ [4]) $ which is now commonly accepted $ [1,2,5].
$

This interpretation implies that the observed correlations have \lq\lq
geometrical\rq\rq\
origin, i.e. they are related to the space-time structure of the volume from
which the pions are emitted $ [6,7]. $
Before this picture is accepted, however, it is important to discuss if there
perhaps exists other dynamical explanations for this phenomenon. In fact, it
was
already a long time ago $ [8] $ pointed out and recalled recently $ [9], $ that
these
effects can be related to another
remarkable  difference between $ \pi^{ \pm} \pi^{ \pm} $
and $ \pi^ +\pi^ - $
channels, namely their resonance structure. While the $ \pi^ +\pi^ - $ system
contains
many resonances, they are absent in the \lq\lq exotic\rq\rq\ $ \pi^{ \pm}
\pi^{ \pm} $ channels.

It is of course an entirely open question if this difference implies or not
a different behaviour of $ \pi \pi $ correlations at very low invariant mass.
However,
there exist an argument suggesting that this may actually be the case $ [8]. $
It
can be summarized as follows.

In the intermediate energy region, the amplitudes describing the $ \pi \pi $
cross-section are known to be dominated by exchange of Regge singularities in
the $ t $-channel. The principle of duality implies that in the exotic channel
the contribution of leading Regge poles exactly cancel each other \footnote{$
^1$}{We
do not consider the Pomeron
 exchange contribution, as it is irrelevant for
short-range correlations we are interested here.}. Consequently, an exotic
channel
is dominated by exchange of singularities with low intercept $ \alpha_ E(0).$
It
follows that the energy-dependent part of the $ \pi^{ \pm} \pi^{ \pm} $
cross-section is expected
to fall with increasing invariant mass of the $ \pi \pi $ system following the
power
law
$$ \sigma \sim \left(M^2 \right)^{\alpha -1}= \left(p_1+p_2 \right)^{2(\alpha
-1)} \eqno (2) $$
with $ \alpha =\alpha_ E(0)\le 0. $

Since there are no resonances in the exotic channel, this smooth behaviour is
expected to hold even at relatively low energies. It is therefore perhaps not
unreasonable to speculate that it continues even down to the
threshold
\footnote{$ ^2 $}{Of course the phase-space factors at the threshold must be
corrected
for. This, however, is to a large extent taken care of, when one considers
{\it normalized} correlation functions $ [8] $.}. This possibility is
particularly
suggestive, when compared to the data of Refs.$ [8,9] $ which show precisely
the same power law \footnote{$^3$}{A closer
 look at the
data $ [1,2] $ shows that they follow a power law in $ Q^2 $ rather than $ M^2.
$ This is not
a serious difficulty: the asymptotic formula (2), justified at large $ M^2 $
cannot distinguish these two possibilities.} at small and medium $ M^2 $ .
In contrast, for non-exotic $ \pi^ +\pi^ - $ channel the leading Regge poles
do
contribute, the intercept $ \alpha =\alpha_ R(0)\simeq 1/2 $ is much higher
and, consequently, the
expected $ M^2 $ dependence much weaker
than in (2). Even more important, the
presence of prominent resonances in the $ s $ channel disturbs
significantly the
simple power-law behaviour and thus precludes
its trivial continuation to the
threshold.

This completes the argument. Admittedly, it is not very convincing as it
requires a rather bold extrapolation. We feel,
however, that it should not be entirely dismissed, particularly in view of the
experimental evidence shown
in $ [8] $ and $ [9]. $ Indeed, if one can (at least
partly) explain the difference between like-change and unlike-change $ \pi \pi
$
correlations without invoking quantum (HBT) interference, the interpretation
of the existing data may be profoundly
 modified. Therefore it seems important to
check carrefully this possibility.

The main point of the present note is to indicate that there exists a feasible
way to test these ideas and perhaps even to give access to a relative strength
of the
effects of quantum (HBT) interference and those of the exotic nature of the
like-charge $ \pi \pi $ interactions.
The idea is there exist particle
combinations which are either

(a)\nobreak\ exotic but not identical

(b)\nobreak\ identical but non-exotic

\nobreak In the category (a) one can list $ K^+\pi^ + $ and $ K^-\pi^ - $
pairs. Since
the particles are
not identical, quantum interference cannot be responsible
for short-range
correlations. Therefore if the effect is present in $ K^{\pm} \pi^{ \pm} $
channels it is
probably due to their \lq\lq exoticity\rq\rq .

\nobreak In the category (b) one has $ \pi^ 0\pi^ 0 $ channel. It is formed by
identical
particles, so if quantum interference is at work
we expect a pattern
similar to $ \pi^ +\pi^ + $ channel. On the other hand $ \pi^ 0\pi^ 0 $ system
forms $ I=2 $ with
probability $ 2/3 $ and $ I=0 $ with probability $ 1/3. $ Only $ I=2 $ state
is exotic and
thus if the resonance structure is responsible we expect a weaker
short-range correlation effect than in $
\pi^ +\pi^ + $
system.

The early data of NA22 coll.$ [10]$ and the more recent data of Aleph
coll. $[11]$ indicate that the very short range correlations  in the
 $K\pi$ exotic
channels, if present at all, are substantially weaker than in the $\pi
\pi$
channel. However, we feel that a precise comparison of the behavior at
very small and at medium $Q^2$ is necessary to assess correctly  the role
of exoticity in the $K\pi$ channel.

To summarize, we suggest that accurate measurements of short-range
correlations in $ K\pi $ and $ \pi^ 0\pi^ 0 $ channels, should help to
determine the origin of
the phenomenon of \lq\lq intermittency\rq\rq . It seems to us important to
clarify this
point before one can safely conclude that the present data can be attributed
solely to the effect of quantum (HBT) interference.
\vskip 24pt
\noindent {\bf ACKNOWLEDGEMENTS}

We would like to thank Wolfram Kittel for the correspondence about data and
       Andrzej
Krzywicki
for interesting comments. This research was supported in part by the
KBN grant $2 0092 91 01.$
\vfill\eject
\centerline{{\bf REFERENCES}}
\vskip 24pt
\item {1} F. Mandl and B. Buschbeck, Vienna-preprint HEPHY-PUB-590-93 (May
1993);
and Proc. XXII Int. Symp. on Multiparticle Dynamics, Santiago de Compostela
Spain, 1992, p.561. A. Pajares
(ed.) World
Scientific, Singapore.

\item {2} N.A. Agababyan et al.,  NA 22 coll.,
{\it Z. Phys.} {\bf C59} (1993) 405.

\item {3} A. Bialas and R. Peschanski, {\it Nucl. Phys.} {\bf B273} (1986) 703.
\item {\nobreak\ } For a recent review, see E.A. DeWolf, I.M. Dremin
  and W. Kittel, to be
published in {\it Phys. Rep.} {\bf C}.

\item {4}
R. Hanbury-Brown and R. Q. Twiss : {\it Phil Mag.} {\bf 45} (1954) 663;
 For a review see D.H. Boal et al., {\it Rev.
Mod. Phys.} {\bf 62} (1990) 553.

\item {5} H. B\"achler et al., NA 35 Coll., {\it Zeit. Phys.} {\bf C61} (1994)
551.

\item {6} P. Grassberger, {\it Nucl. Phys.} {\bf B120} (1977) 231.

\item {7} A. Bialas, {\it Acta Phys. Pol.} {\bf B23} (1992) 561.

\item {8} E.L. Berger, R. Singer, G.M. Thomas and T. Kafka,
{\it Phys. Rev.} {\bf D15}
(1977) 206.

\item {9} I.V. Ajinenko et al., NA22 Coll., {\it Zeit. Phys.}  {\bf C61} (1994)
567.

\item {10} M. Adamus   et al., NA22 Coll., {\it Zeit. Phys.}  {\bf C37} (1988)
347, fig.3 e,f.

\item {11} D. Decamp et al., NA22 Coll., {\it Zeit. Phys.}  {\bf C54} (1992)
75, fig.7.

\end